\documentclass[conference]{IEEEtran}
\IEEEoverridecommandlockouts
\usepackage[a4paper,width=170mm,top=20mm,bottom=26mm]{geometry}
\usepackage{amsmath,amssymb,amsfonts}
\usepackage{algorithmic}
\usepackage{graphicx}
  \graphicspath{{figs/}}
\usepackage{textcomp}
\usepackage[dvipsnames,table,xcdraw]{xcolor}
\usepackage[frozencache,cachedir=minted-cache]{minted}
\usepackage[capitalize]{cleveref}
\usepackage{url}
\usepackage[xindy, nonumberlist]{glossaries}
\usepackage[font=small]{caption}
\usepackage{subcaption}
\usepackage{siunitx}
\usepackage{booktabs}
\usepackage[super]{nth}
\usepackage[inline]{enumitem}
\usepackage{comment}
\makenoidxglossaries%

\newacronym{SigMF}{SigMF}{Signal Metadata Format}

\newacronym{2d}{2D}{two-dimensional}
\newacronym{3d}{3D}{three-dimensional}
\newacronym{3gpp}{3GPP}{3rd generation partnership project}
\newacronym{5g}{5G}{fifth-generation}
\newacronym{6g}{6G}{sixth-generation}
\newacronym{api}{API}{application program interface}
\newacronym{abp}{ABP}{authentication by personalisation}
\newacronym{aclr}{ACLR}{adjacent channel leakage ratio}
\newacronym{adc}{ADC}{Analog-to-digital converter}
\newacronym{adr}{ADR}{adaptive data rate}
\newacronym{ag}{AG}{array gain}
\newacronym{ai}{AI}{artificial intelligence}
\newacronym{amam}{AM/AM}{amplitude modulation to amplitude modulation}
\newacronym{amp}{AMP}{approximate message passing}
\newacronym{ampm}{AM/PM}{amplitude modulation to phase modulation}
\newacronym{aoa}{AOA}{angle-of-arrival}
\newacronym{aod}{AOD}{angle-of-departure}
\newacronym{ap}{AP}{access point}
\newacronym{apu}{APU}{access point unit}
\newacronym{ar}{AR}{augmented reality}
\newacronym{arp}{ARP}{Antenna Reference Point}
\newacronym{asic}{ASIC}{application specific integrated circuit}
\newacronym{ask}{ASK}{amplitude-shift keying}
\newacronym{auv}{AUV}{autonomous underwater vehicle}
\newacronym{awgn}{AWGN}{additive white Gaussian noise}
\newacronym{baw}{BAW}{bulk acoustic wave}
\newacronym{bb}{BB}{base-band}
\newacronym{bf}{BF}{beamforming}
\newacronym{bldc}{BLDC}{brushless DC}
\newacronym{bom}{BOM}{bill of materials}
\newacronym{bs}{BS}{base station}
\newacronym{bw}{BW}{bandwidth}
\newacronym{cad}{CAD}{channel activity detection}
\newacronym{cars}{CARS}{calibration reference signal}
\newacronym{cbm}{CBM}{condition based maintenance}
\newacronym{cc}{CC}{constant current}
\newacronym{ccdf}{CCDF}{complementary cumulative distribution function}
\newacronym{ccnn}{CCNN}{circular convolutional neural network}
\newacronym{ccs}{CCS}{correlative channel sounder}
\newacronym{cdf}{CDF}{cumulative distribution function}
\newacronym{cdrx}{CDRX}{connected mode DRX}
\newacronym{ce}{CE}{coverage enhancement}
\newacronym{ced}{CED}{cumulative energy density}
\newacronym{cf}{CF}{cell-free}
\newacronym{cfo}{CFO}{carrier frequency offset}
\newacronym{cir}{CIR}{channel impulse response}
\newacronym{cost}{COST}{commercial off-the-shelf}
\newacronym{cots}{COTS}{commercial off-the-shelf}
\newacronym{cp}{CP}{cyclic prefix}
\newacronym{cpt}{CPT}{capacitive power transfer}
\newacronym{cpu}{CPU}{central-processing unit}
\newacronym{cqi}{CQI}{channel quality indicator}
\newacronym{cr}{CR}{coding rate}
\newacronym{crlb}{CRLB}{Cram\'er-Rao lower bound}
\newacronym{crs}{CRS}{cell reference signal}
\newacronym{cs}{CS}{compressed sensing}
\newacronym{csi}{CSI}{channel state information}
\newacronym{csp}{CSP}{contact service point}
\newacronym{css}{CSS}{chirp spread spectrum}
\newacronym{cv}{CV}{constant voltage}
\newacronym{dac}{DAC}{digital-to-analog converter}
\newacronym{daq}{DAQ}{data acquisition system}
\newacronym{dc}{DC}{direct current}
\newacronym{dcc}{DCC}{dynamic cooperation clustering}
\newacronym{de}{DE}{drain efficiency}
\newacronym{dl}{DL}{downlink}
\newacronym{dlc}{DLC}{Distributed Laser Charging}
\newacronym{dma}{DMA}{Direct Memory Access}
\newacronym{dmimo}{D-MIMO}{distributed MIMO}
\newacronym{dpd}{DPD}{digital pre-distortion}
\newacronym{dpdk}{DPDK}{Data Plane Development Kit}
\newacronym{drx}{DRX}{Discontinuous Reception Mode}
\newacronym{duc}{DUC}{digital up-converter}
\newacronym{dss}{DSS}{Dataset Storage Standard}
\newacronym{esr}{ESR}{equivalent series resistance}
\newacronym{e2e}{E2E}{end-to-end}
\newacronym{easa}{EASA}{European Union Aviation Safety Agency}
\newacronym{ecdf}{eCDF}{empirical cumulative distribution function}
\newacronym{ecsp}{ECSP}{edge computing service point}
\newacronym{edlc}{EDLC}{electrostatic double-layer capacitors}
\newacronym{edrx}{eDRX}{Extended Discontinuous Reception Mode}
\newacronym{ee}{EE}{energy efficiency}
\newacronym{egprs}{EGPRS}{Enhanced Data Rates for GSM Evolution}
\newacronym{eirp}{EIRP}{effective isotropic radiated power}
\newacronym{em}{EM}{electromagnetic}
\newacronym{embb}{eMBB}{enhanced Mobile Broadband}
\newacronym{en}{EN}{energy neutral}
\newacronym{end}{END}{energy-neutral device}
\newacronym{eol}{EoL}{end of life}
\newacronym{epu}{EPU}{edge processing unit}
\newacronym{erp}{ERP}{effective radiated power}
\newacronym[plural=ESCs,firstplural=electronic speed controllers (ESCs)]{esc}{ESC}{electronic speed control}
\newacronym{etsi}{ETSI}{European Telecommunications Standards Institute}
\newacronym{evm}{EVM}{Error Vector Magnitude}
\newacronym{fair}{FAIR}{Findability, Accessibility, Interoperability, and Reuse of digital assets}
\newacronym{fa}{FA}{federation anchor}
\newacronym{fd}{FD}{front-haul distance}
\newacronym{fembb}{feMBB}{further enhanced mobile broadband}
\newacronym{fft}{FFT}{fast Fourier transform}
\newacronym{fh}{FH}{fronthaul}
\newacronym{fim}{FIM}{Fisher information matrix}
\newacronym{fom}{FoM}{figure of merit}
\newacronym{fpga}{FPGA}{field-programmable gate array}
\newacronym{gb}{GB}{grant-based}
\newacronym{gf}{GF}{grant-free}
\newacronym{gnb}{gNB}{Next Generation Node B}
\newacronym{gnn}{GNN}{graph neural network}
\newacronym{gnss}{GNSS}{global navigation satellite system}
\newacronym{gpclk}{GPCLK}{general purpose clock}
\newacronym{gprs}{GPRS}{General Packet Radio Services}
\newacronym{gps}{GPS}{Global Positioning System}
\newacronym{gpu}{GPU}{graphical processing unit}
\newacronym{gsm}{GSM}{Global System for Mobile Communications}
\newacronym{gwp}{GWP}{Global Warming Potential}
\newacronym{harq}{HARQ}{hybrid automatic repeat request}
\newacronym{hat}{HAT}{hardware attached on top}
\newacronym{hcs}{HCS}{human-centric services}
\newacronym{hpbm}{HPBM}{half power beam width}
\newacronym{HyMPRo}{HyMPRo}{Hybrid Multi-Path Routing algorithm}
\newacronym{ib}{IB}{in-band}
\newacronym{ibo}{IBO}{input back-off}
\newacronym{ic}{IC}{integrated circuit}
\newacronym{id}{ID}{information decoding}
\newacronym{if}{IF}{intermediate-frequency}
\newacronym{iid}{i.i.d.}{independently and identically distributed}
\newacronym{im}{IM}{intermodulation}
\newacronym{imu}{IMU}{inertial measurement unit}
\newacronym{io}{IO}{input/output}
\newacronym{iot}{IoT}{Internet of Things}
\newacronym{ipt}{IPT}{inductive power transfer}
\newacronym{ipy}{IPY}{Interventions per Year}
\newacronym{iq}{IQ}{in-phase and quadrature}
\newacronym{ir}{IR}{infrared}
\newacronym{ism}{ISM}{industrial, scientific and medical}
\newacronym{isp}{ISP}{internet service provider}
\newacronym{kpi}{KPI}{key performance indicator}
\newacronym{kvi}{KVI}{key value indicator}
\newacronym{lna}{LNA}{low-noise amplifier}
\newacronym{larva}{LARVA}{LARge Virtual Array}
\newacronym{lca}{LCA}{life cycle assessment}
\newacronym{lco}{LCO}{lithium cobalt oxide}
\newacronym{ldo}{LDO}{Low-dropout voltage regulator}
\newacronym{ldpc}{LDPC}{low-density parity-check}
\newacronym{lfp}{LFP}{lithium iron phosphate}
\newacronym{lic}{LIC}{lithium-ion capacitor}
\newacronym{lidar}{LiDAR}{light detection and ranging}
\newacronym{liion}{Li-ion}{lithium-ion}
\newacronym{lipo}{LiPo}{lithium polymer}
\newacronym{lis}{LIS}{large intelligent surface}
\newacronym{llh}{LLH}{log-likelihood}
\newacronym{lmmse}{LMMSE}{least minimum mean square error}
\newacronym{lmo}{LMO}{lithium ion manganese oxide}
\newacronym{lo}{LO}{local oscillator}
\newacronym{lora}{LoRa}{long range}
\newacronym{lorawan}{LoRaWAN}{long-range wide-area network}
\newacronym{los}{LoS}{line-of-sight}
\newacronym{lp}{LP}{linear programming}
\newacronym{lpt}{LPT}{laser power transfer}
\newacronym{lpwa}{LPWA}{Low Power Wide Area}
\newacronym{lpwan}{LPWAN}{low-power wide-area network}
\newacronym{lqi}{LQI}{link quality indicator}
\newacronym{lrelu}{LReLU}{leaky rectified linear unit}
\newacronym{lrt}{LRT}{likelihood-ratio test}
\newacronym{ls}{LS}{least squares}
\newacronym{lsa}{LSA}{large synthetic array}
\newacronym{lsfc}{LSFC}{large-scale fading component}
\newacronym{ltc}{LTC}{lithium thionyl chloride}
\newacronym{lte}{LTE}{Long Term Evolution}
\newacronym{lto}{LTO}{lithium titanate}
\newacronym{m2m}{M2M}{machine to machine}
\newacronym{mac}{MAC}{Medium Access Control}
\newacronym{mate}{MATE}{millimeter-wave MIMO testbed}
\newacronym{mcl}{MCL}{Maximum Coupling Loss}
\newacronym{mcs}{MCS}{modulation and coding scheme}
\newacronym{mcu}{MCU}{Microcontroller Unit}
\newacronym{mec}{MEC}{multi-access edge computing}
\newacronym{mems}{MEMS}{micro-electromechanical systems}
\newacronym{mf}{MF}{matched filter}
\newacronym{mimo}{MIMO}{multiple-input multiple-output}
\newacronym{miso}{MISO}{multiple-input single-output}
\newacronym{ml}{ML}{machine learning}
\newacronym{mlp}{MLP}{multilayer perceptron}
\newacronym{mmimo}{mMIMO}{massive MIMO}
\newacronym{mmse}{MMSE}{minimum mean square error}
\newacronym{mmtc}{mMTC}{massive machine-typed communication}
\newacronym{mmwave}{mmWave}{millimeter wave}
\newacronym{mpc}{MPC}{multipath component}
\newacronym{mppt}{MPPT}{maximum power point tracking}
\newacronym{mr}{MR}{maximum ratio}
\newacronym{mrc}{MRC}{maximum ratio combining}
\newacronym{mrc_em}{MRC}{maximum ratio combining}
\newacronym{mrc_EM}{MRC}{Magnetic Resonance Coupling}
\newacronym{mrt}{MRT}{maximum ratio transmission}
\newacronym{mse}{MSE}{mean square error}
\newacronym{mtc}{MTC}{Machine-Type Communication}
\newacronym{multi-rat}{Multi-RAT}{multiple radio access technology}
\newacronym{nb}{NB}{narrowband}
\newacronym{nbiot}{NB-IoT}{narrowband IoT}
\newacronym{nca}{NCA}{nickel cobalt aluminum}
\newacronym{nicd}{NiCd}{nikkel cadmium}
\newacronym{nimh}{NiMH}{nikkel metal hydride}
\newacronym{nlos}{NLoS}{non-line-of-sight}
\newacronym{nmc}{NMC}{nickel manganese cobalt}
\newacronym{nn}{NN}{neural network}
\newacronym{nnls}{NNLS}{non-negative least squares}
\newacronym{noma}{NOMA}{non-orthogonal multiple access}
\newacronym{nr}{NR}{New Radio}
\newacronym{ntp}{NTP}{network time protocol}
\newacronym{ni}{NI}{National Instruments}
\newacronym{netcdf}{NetCDF}{Network Common Data Form}
\newacronym{oai}{OAI}{OpenAirInterface} %
\newacronym{ofdm}{OFDM}{orthogonal frequency-division multiplexing}
\newacronym{ofdmim}{OFDM-IM}{OFDM with index modulation}
\newacronym{oob}{OOB}{out-of-band}
\newacronym{oran}{O-RAN}{open radio-access network}
\newacronym{os}{OS}{operating system}
\newacronym{ota}{OTA}{over-the-air}
\newacronym{otaa}{OTAA}{over-the-air authentication}
\newacronym{p2p}{P2P}{point-to-point}
\newacronym{pa}{PA}{power amplifier}
\newacronym{pae}{PAE}{power-added efficiency}
\newacronym{papr}{PAPR}{peak-to-average power ratio}
\newacronym{pc}{PC}{pilot count}
\newacronym{pcb}{PCB}{printed circuit board}
\newacronym{pcg}{PCG}{power consumption gain}
\newacronym{pd}{PD}{powered device}
\newacronym{phy}{PHY}{physical}
\newacronym{pdcch}{PDCCH}{physical downlink control channel}
\newacronym{pdp}{PDP}{power delay profile}
\newacronym{pdsch}{PDSCH}{physical downlink shared channel}
\newacronym{peb}{PEB}{positioning error bound}
\newacronym{per}{PER}{packet error rate}
\newacronym{pg}{PG}{path gain}
\newacronym{pgd}{PGD}{proximal gradient descent}
\newacronym{pl}{PL}{path loss}
\newacronym{pll}{PLL}{phase-locked loop}
\newacronym{poe}{PoE}{power-over-Ethernet}
\newacronym{pps}{1PPS}{pulse per second}
\newacronym{prbs}{PRBs}{Physical Resource Blocks}
\newacronym{ps}{PS}{Processing System}
\newacronym{psd}{PSD}{power spectral density}
\newacronym{pse}{PSE}{power sourcing equipment}
\newacronym{psm}{PSM}{power saving mode}
\newacronym{pss}{PSS}{primary synchronisation signal}
\newacronym{ptp}{PTP}{precision-time protocol}
\newacronym{ptrs}{PTRS}{Phase-Tracking Reference Signals}
\newacronym{ptw}{PTW}{paging time window}
\newacronym{pwm}{PWM}{pulse width modulation}
\newacronym{qam}{QAM}{quadrature amplitude modulation}
\newacronym{qos}{QoS}{quality-of-service}
\newacronym{quadriga}{QuaDRiGa}{QUAsi Deterministic RadIo channel GenerAtor}
\newacronym{ra}{RA}{Random Access}
\newacronym{ran}{RAN}{radio access network}
\newacronym{rar}{RAR}{Random Access Response}
\newacronym{rat}{RAT}{radio access technology}
\newacronym{rb}{Rb}{Rubidium}
\newacronym{rbs}{RBS}{radio base station}
\newacronym{re}{RE}{radio element}
\newacronym[]{relu}{ReLU}{rectified linear unit}
\newacronym{rf}{RF}{radio frequency}
\newacronym{rfeh}{RFEH}{radio frequency energy harvesting}
\newacronym{rfic}{RFIC}{radio-frequency integrated circuit}
\newacronym{rfid}{RFID}{radio frequency identification}
\newacronym{rfpt}{RFPT}{radio frequency power transfer}
\newacronym{rfsoc}{RFSoC}{Radio Frequency System-on-Chip}
\newacronym{ris}{RIS}{reflective intelligent surface}%
\newacronym{rllmtc}{RLLMTC}{reliable low latency machine type communication}
\newacronym{rls}{RLS}{recursive least squares}
\newacronym{rms}{RMS}{root-mean-square}
\newacronym{rmse}{RMSE}{root-mean-square error}
\newacronym{rof}{RoF}{radio-over-fiber}
\newacronym{ros}{ROS}{robot operating system}
\newacronym{rpi}{RPi}{Raspberry Pi}
\newacronym{rrc}{RRC}{Radio Resource Connection}
\newacronym{rreq}{RREQ}{route request packet}
\newacronym{rsrp}{RSRP}{Reference Signals Received Power}
\newacronym{rss}{RSS}{received signal strength}
\newacronym{rssi}{RSSI}{received signal strength indicator}
\newacronym{rtc}{RTC}{real time clock}
\newacronym{rtk}{RTK}{real time kinematics}
\newacronym{rts}{RTS}{ray tracing simulator}
\newacronym{rw}{RW}{RadioWeaves}
\newacronym{rx}{RX}{receiver}
\newacronym{rzf}{RZF}{regularized zero forcing}
\newacronym{rir}{RIR}{room impulse response}
\newacronym{s-parameter}{S-parameter}{scattering parameter}
\newacronym{sa}{SA}{synchronization anchor}
\newacronym{scfdma}{SCFDMA}{single-carrier frequency division multiple access}
\newacronym{sdg}{SDG}{Sustainable Development Goal}
\newacronym{sdm}{SDM}{sigma-delta modulator}
\newacronym{sdn}{SDN}{software-defined network}
\newacronym{sdof}{SDoF}{sigma-delta over fiber}
\newacronym{sdr}{SDR}{software-defined radio}
\newacronym{sf}{SF}{spreading factor}
\newacronym{sfn}{SFN}{single frequency network}
\newacronym{sfp}{SFP}{small form-factor pluggable}
\newacronym{sinr}{SINR}{signal-to-interference-plus-noise ratio}
\newacronym{siso}{SISO}{single-input single-output}
\newacronym{slam}{SLAM}{simultaneous localization and mapping}
\newacronym{slc}{SLC}{spatial leakage suppression}
\newacronym{smc}{SMC}{specular multipath component}
\newacronym{smps}{SMPS}{switched mode power supply}
\newacronym{sndr}{SNDR}{signal-to-noise-and-distortion ratio}
\newacronym{snidr}{SNIDR}{signal-to-noise-and-interference-and-distortion ratio}
\newacronym{snr}{SNR}{signal-to-noise ratio}
\newacronym{soc}{SoC}{state of charge}
\newacronym{ssb}{SSB}{synchronisation signal block}
\newacronym{ssd}{SSD}{solid state drive}
\newacronym{steam}{STEAM}{science, technology, engineering, the arts, and mathematics}
\newacronym{svd}{SVD}{singular value decomposition}
\newacronym{tau}{TAU}{tracking area update}
\newacronym{tcer}{TCER}{transported to consumed energy ratio}
\newacronym{tdd}{TDD}{time division duplexing}
\newacronym{tdoa}{TDOA}{time-difference-of-arrival}
\newacronym{toa}{TOA}{time-of-arrival}
\newacronym{tof}{ToF}{time-of-flight}
\newacronym{tosm}{TOSM}{through-open-short-match}
\newacronym{trl}{TRL}{technology readyness level}
\newacronym{trp}{TRP}{Transmission Reception Point}
\newacronym{tsn}{TSN}{time-sensitive networking}
\newacronym{ttm}{TTM}{time to market}
\newacronym{ttn}{TTN}{The Things Network}
\newacronym{tx}{TX}{transmitter}
\newacronym{tcp}{TCP}{Transmission Control Protocol}
\newacronym{uav}{UAV}{unmanned aerial vehicle}
\newacronym{udp}{UDP}{User Datagram Protocol}
\newacronym{ue}{UE}{user equipment}
\newacronym{ugv}{UGV}{unmanned ground vehicle}
\newacronym{uhd}{UHD}{USRP hardware driver}
\newacronym{uhf}{UHF}{ultra-high frequency}
\newacronym{ul}{UL}{uplink}
\newacronym{ula}{ULA}{uniform linear array}
\newacronym{ummtc}{umMTC}{ultra massive machine type communication}
\newacronym{upa}{UPA}{uniform planar array}
\newacronym{ura}{URA}{uniform rectangular array}
\newacronym{urllc}{URLLC}{ultra-reliable low-latency communications}
\newacronym{usrp}{USRP}{universal software radio peripheral}
\newacronym{uv}{UV}{unmanned vehicle}
\newacronym{uwb}{UWB}{ultrawideband}
\newacronym{vep}{VEP}{virtual edge platform}
\newacronym{vlc}{VLC}{visible light communication}
\newacronym{vlp}{VLP}{visible light positioning}
\newacronym{vna}{VNA}{vector network analyzer}
\newacronym{vr}{VR}{virtual reality}
\newacronym{wb}{WB}{wideband}
\newacronym{wpt}{WPT}{wireless power transfer}
\newacronym{wr}{WR}{White Rabbit}
\newacronym{wrsn}{WRSN}{wireless rechargeable sensor network}
\newacronym{wsn}{WSN}{wireless sensor network}
\newacronym{xr}{XR}{extended reality}
\newacronym{z3ro}{Z3RO}{zero third-order distortion}
\newacronym{zf}{ZF}{zero-forcing}
\newacronym{zmcscg}{ZMCSCG}{zero mean circularly symmetric complex Gaussian}
\newacronym{hdf5}{HDF5}{Hierarchical Data Format version 5}

\newacronym{fits}{FITS}{Flexible Image Transport System}
\newacronym{votable}{VOTable}{Virtual Observatory Table}%
\def\BibTeX{{\rm B\kern-.05em{\sc i\kern-.025em b}\kern-.08em
    T\kern-.1667em\lower.7ex\hbox{E}\kern-.125emX}}

\usepackage[backend=biber,style=ieee]{biblatex}
\AtBeginBibliography{\footnotesize}

\usepackage{tikz}
\usepackage{pgfplots}
\usepgfplotslibrary{external} 
\usetikzlibrary{external}
\pgfplotsset{every axis plot/.append style={thick}}
\tikzset{external/mode=graphics if exists}

\pgfplotsset{
ylabel style={align=center},
xlabel style={yshift=1mm},
grid style=dotted,
axis lines* = {left},
legend style={fill opacity=0.8, draw opacity=1, text opacity=1, draw=white!80.00000!black, at={(0.3,0.45)}},
legend style={nodes={scale=0.9, transform shape}},
legend style={draw=none},
tick align=outside,
tick pos=left,
axis background/.style={fill=white},
axis x line*=bottom,
axis y line*=left,
width = 0.9\linewidth,
height = 0.556\linewidth,
x grid style={white!69.01961!black},
xtick style={color=black},
y grid style={white!69.01961!black},
ytick style={color=black},
enlargelimits=false,
xmajorgrids,
ymajorgrids,
yticklabel style = {font=\footnotesize,xshift=0.5ex},
xticklabel style = {font=\footnotesize,yshift=0.5ex},
ylabel style = {font=\small},
xlabel style = {font=\small}
}
\newcommand{\tikzxmark}{%
\tikz[scale=0.23] {
    \draw[line width=0.7,line cap=round] (0,0) to [bend left=6] (1,1);
    \draw[line width=0.7,line cap=round] (0.2,0.95) to [bend right=3] (0.8,0.05);
}}
\newcommand{\tikzcmark}{%
\tikz[scale=0.23] {
    \draw[line width=0.7,line cap=round] (0.25,0) to [bend left=10] (1,1);
    \draw[line width=0.8,line cap=round] (0,0.35) to [bend right=1] (0.23,0);
}}
\newcommand{\yes}{\protect\tikzcmark}
\newcommand{\no}{\protect\tikzxmark}

\usepackage{array}
\newcolumntype{H}{>{\setbox0=\hbox\bgroup}c<{\egroup}@{}}

\newlength{\figurewidth}
\newlength{\figureheight}

\begin{document}

\title{An Open Dataset Storage Standard\\ for 6G Testbeds
\thanks{6GTandem has received funding from the Smart Networks and Services Joint Undertaking (SNS JU) under the European Union's Horizon Europe research and innovation programme under Grant Agreement No~101096302. The REINDEER project has received funding from the European Union's Horizon 2020 research and innovation programme under grant agreement No.~101013425.}
}

\author{
    \IEEEauthorblockN{
    Gilles Callebaut\IEEEauthorrefmark{1}, 
    Michiel Sandra\IEEEauthorrefmark{2},
    Christian Nelson\IEEEauthorrefmark{2},
    Thomas Wilding\IEEEauthorrefmark{3},\\
    Daan Delabie\IEEEauthorrefmark{1},
    Benjamin J.\,B. Deutschmann\IEEEauthorrefmark{3},
    William T\"{a}rneberg\IEEEauthorrefmark{2},\\
    Emma Fitzgerald\IEEEauthorrefmark{2},
    Anders J. Johansson\IEEEauthorrefmark{2},
    Liesbet Van der Perre\IEEEauthorrefmark{1}
}
\IEEEauthorblockA{\IEEEauthorrefmark{1}
        KU Leuven, ESAT-Wavecore, Ghent Technology Campus, B-9000 Ghent, Belgium
}
\IEEEauthorblockA{\IEEEauthorrefmark{2}
        Department of Electrical and Information Technology, Lund University, SE-221 00 Lund, Sweden
    }%
\IEEEauthorblockA{\IEEEauthorrefmark{3}
    Graz University of Technology, Austria
      }%
}

\maketitle

\begin{abstract}
    The emergence of sixth-generation (6G) networks has spurred the development of novel testbeds, including sub-THz networks, cell-free systems, and 6G simulators. To maximize the benefits of these systems, it is crucial to make the generated data publicly available and easily reusable by others. Although data sharing has become a common practice, a lack of standardization hinders data accessibility and interoperability. In this study, we propose the Dataset Storage Standard (DSS) to address these challenges by facilitating data exchange and enabling convenient processing script creation in a testbed-agnostic manner. DSS supports both experimental and simulated data, allowing researchers to employ the same processing scripts and tools across different datasets. Unlike existing standardization efforts such as SigMF and NI RF Data Recording API, DSS provides a broader scope by accommodating a common definition file for testbeds and is not limited to RF data storage. The dataset format utilizes a hierarchical structure, with a tensor representation for specific experiment scenarios. In summary, DSS offers a comprehensive and flexible framework for enhancing the FAIR principles (Findability, Accessibility, Interoperability, and Reusability) in 6G testbeds, promoting open and efficient data sharing in the research community.
\end{abstract}

\begin{IEEEkeywords}
6G, dataset, measurements, testbed, channel sounding, simulations
\end{IEEEkeywords}

\glsresetall
\section{Introduction}
Both academia and industry are actively developing new \gls{6g} testbeds, such as sub-THz networks, \gls{cf} systems, and \gls{6g} simulators. 
To benefit from these systems, the measurements, or more generally, the generated data should be publicly available and convenient to be reused by others. 
While the former has become the norm, the latter has not. 
Therefore, a common data format and interface are imperative to adhere to the \gls{fair} principles~\cite{wilkinson2016fair}. 

We propose a \gls{dss}~\cite{dss2023} i) to facilitate exchanging data and ii) to conveniently create processing scripts in a testbed-agnostic fashion. 
The latter enables developers and researchers to share processing scripts for the testbeds without intermediate scripts to tailor it to the specific output format of one testbed.
Additionally, \gls{dss} is designed to also store data originating from simulations, enabling the processing of experimental and simulation with the same scripts and tools. 
The open-source interface documentation, including examples, is hosted on GitHub. This is a living document; for the most up-to-date information, please refer to the GitHub repository~\cite{dss2023}. 

As evident from~\cite[Table~XI and~XII]{OnTheRoadTo6G} and~\cite{callebaut20236g}, \emph{a significant number of testbeds are presently in the design and construction stages for 6G. This underscores the pressing need for standardized storage, processing, and documentation of experiments. Fortunately, \gls{dss} arrives at precisely the right moment to address this requirement.}

\begin{table*}[hbtp]
\centering
\caption{Comparison of Data Storage Standards}\label{tab:comparison}
\resizebox{0.95\textwidth}{!}{%
\begin{tabular}{llllHl}
\toprule
\textbf{Standard} & \textbf{Data Stored} & \textbf{Storage Method} & \textbf{Focus} & \textbf{Supported Languages} & \textbf{Hardware description} \\ \hline 
NetCDF              & Structured scientific data    & Hierarchical format       & General-purpose & C, C++, Fortran, Python & \no \\ 
HDF5                & Structured and unstructured data & Hierarchical format  & General-purpose & C, C++, Fortran, Python & \no \\ 
 Digital RF          & RF and signal data             & Custom format          & RF signal data  & C, Python             & \no \\ 
SigMF               & Signal metadata                & Lightweight format      & Signal metadata  & Any (Text-based)       & \no \\ 
NI RF Data API       & RF data (NI-specific)                        & SigMF format                   & RF data management & C, C++, Python       & \yes (only NI) \\ 
\Acrshort{fits}       & Measurement-specific                        & Table structures                   & Astronomical observations  & Python, Matlab, YAML       & \no \\
\Acrshort{votable}       & Measurement-specific                        & Hierarchical tabular structure                   &  Astronomical observations  & Python, Matlab, YAML       & \no \\
\rowcolor{gray!10}
\textbf{\Acrshort{dss} }      & Measurement-specific                        & Custom format (HDF5 + NetCDF)                   & 6G experiments and data storage  & Python, Matlab, YAML       & \yes \\\bottomrule
\end{tabular}%
}
\end{table*}

Several standards are being proposed and used in both industry and academia to record or store measurement data. Examples are \gls{netcdf}, \gls{hdf5}, Digital~RF~\cite{DigitalRF}, \gls{SigMF}~\cite{sigMF}, and NI RF Data Recording API~\cite{NI-RF}. \Gls{netcdf} and \gls{hdf5} are general-purpose data formats used in various scientific domains, while Digital~RF and \gls{SigMF} are specialized formats for \gls{rf} and signal data. The NI RF Data Recording API is specific to National Instruments hardware and software solutions. Next to \gls{rf} research, several data storage standards are proposed in astronomy, having similar requirements~\cite{SurveyStandardsAstronomy, gunst2018antenna} to \gls{dss}. \Gls{fits} is one of the most widely used data storage standards in astronomy. It is a flexible and self-descriptive format, primarily designed for storing astronomical images and tables. \Gls{fits} files include headers that contain metadata describing the content of the data. The format is highly extensible and supports various data types, making it suitable for a wide range of astronomical data. Another standard is \gls{votable}, which is an XML-based standard designed for representing tabular data, such as catalogs and data tables. It is widely used in the Virtual Observatory framework, facilitating the exchange of data between different observatories and data archives. \gls{votable} files include metadata and descriptions of table columns, making them self-descriptive and interoperable. Although having some similarities in the requirements, \gls{fits} and \gls{votable} are tailored to astronomy and are therefore not directly applicable to 6G testbeds. An overview and differences of these standards are provided in~\cref{tab:comparison}.

In contrast to other standardization endeavors in the context of 6G testbeds, such as \gls{SigMF} and NI RF Data Recording API, \gls{dss} has a broader scope. It supports a common definition file for testbeds and is not limited to storing RF data exclusively. \Gls{dss} emphasizes describing hardware-agnostic experiment descriptions. While not designed for experiment configuration, \gls{dss} does not impose limitations on the use of it for that purpose. Furthermore, \gls{dss} supports mobility, enabling the movement of the \glspl{ue} about the testbed. This can also be very useful for creating a database of different experiments.

At its core, \gls{dss} is built upon the data model of the widely recognized \gls{hdf5} and \gls{netcdf} storage standards, and data can be stored using either of these formats. \Gls{dss} introduces an additional layer in the form of metadata description files, customizing it to better suit the requirements of 6G testbeds.

\begin{figure*}[hbtp]
  \centering
    \includegraphics[width=0.95\linewidth]{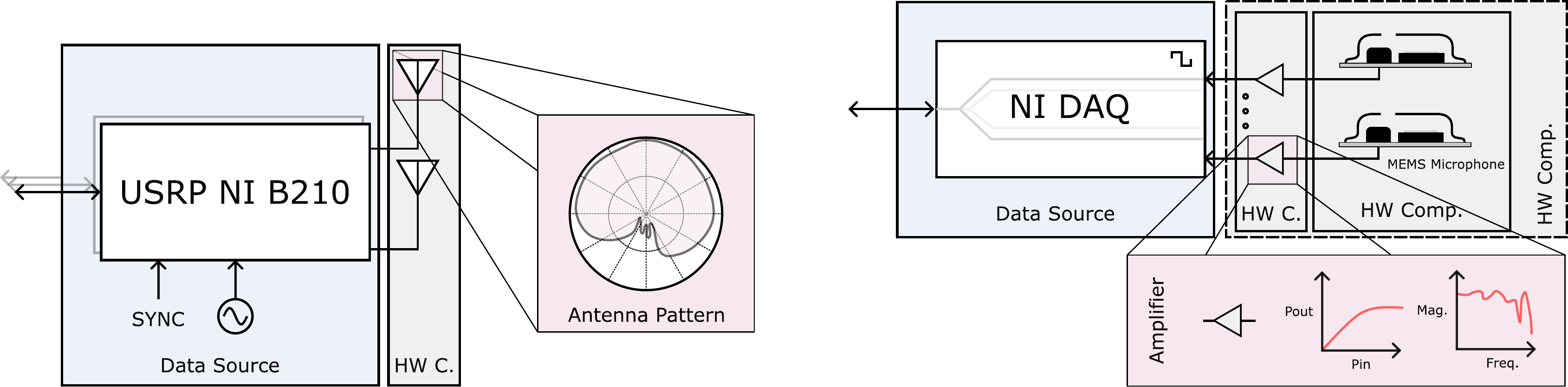}
    \caption{Example of two testbed setups (inspired by the Techtile testbed~\cite{9815696}). \textbf{Left}: Illustrating multiple \gls{usrp} NI B210 \glspl{sdr} having each two antennas connected. The antenna pattern is an attribute of the specific antenna hardware component. \textbf{Right}: One central \gls{daq} samples multiple \gls{mems} microphones. Each microphone is connected to an amplifier before being sampled. Two examples of attributes related to an amplifier is depicted, i.e., the \gls{amam} and frequency response.}%
    \label{fig:testbed-example}
\end{figure*}

\section{DSS Architecture}\label{sec:architecture}
\begin{figure*}
  \centering
      \includegraphics[width=0.95\linewidth]{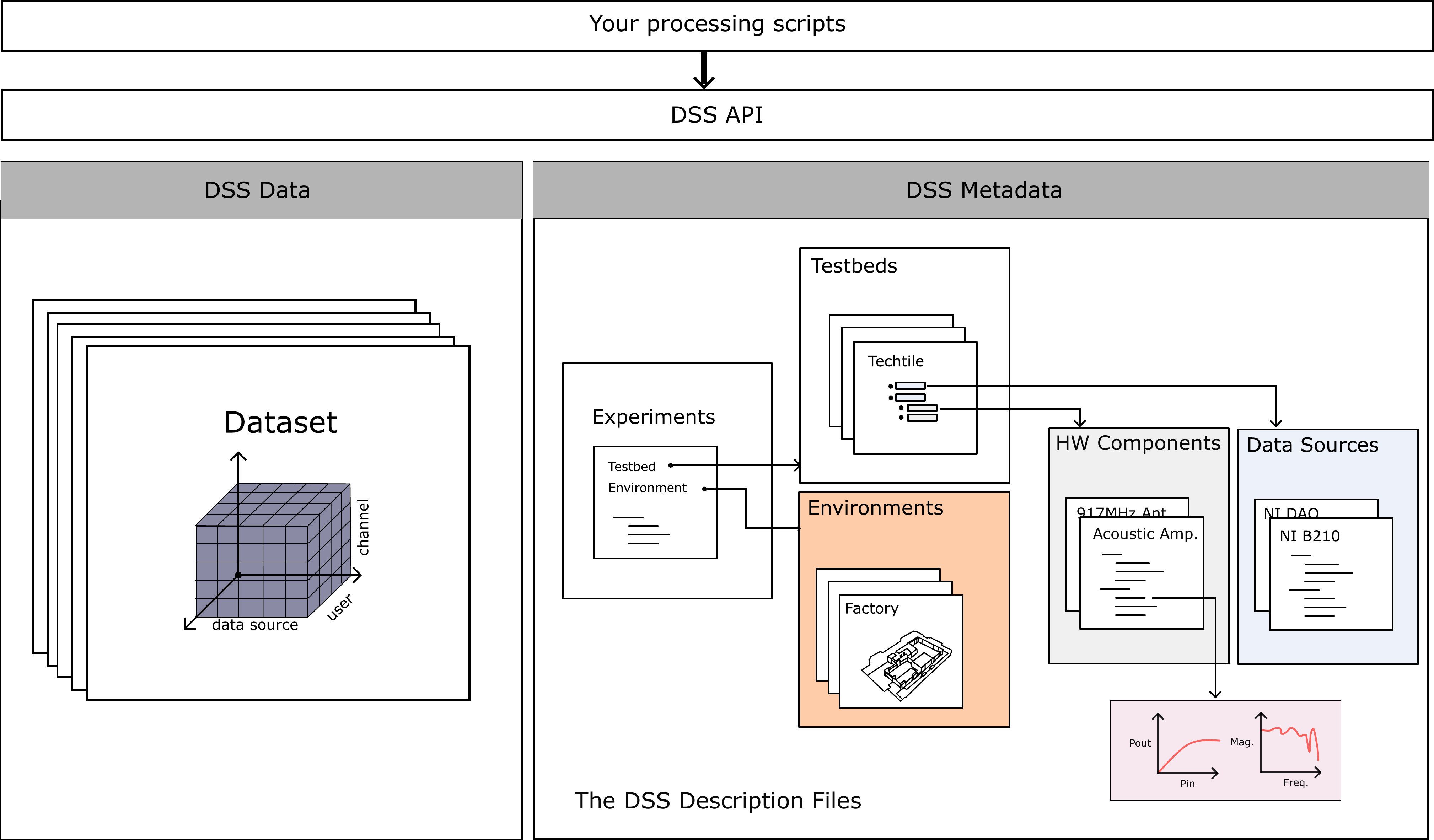}
    \caption{Illustration of the proposed standard, consisting of the description files, datasets, and API.}%
    \label{fig:overiew-dss}
\end{figure*}

The standard comprises three main components:
i) human-readable description files detailing experiments and testbeds (\cref{sec:description-files}), ii) a specific data storage format based on the type of measurement (\cref{sec:dataset}), iii) and an \gls{api} (\cref{sec:dataset}).
These are illustrated in~\cref{fig:overiew-dss} and elaborated below.

The \emph{description files} are used to i) programmatically extract relevant information regarding the stored dataset, and ii) to model the hardware setup to perform simulations. Consequently, the behavior of hardware is defined in the \emph{hardware component}. This can range from the frequency response of the utilized cables to the \gls{pa} \gls{amam} profile. By defining how to store and describe this metadata, the \gls{fair} principles are forced to be met and calibration procedures and hardware imperfections become transparent to the users of the testbed and the dataset.

\subsection{The DSS Description Files}\label{sec:description-files}
A number of files are required to interpret and explain a conducted experiment. 
The files are structured in a way such that they can conveniently be re-used in different experiments and testbeds. 
For instance, a \emph{data source} can describe an \gls{sdr}, as well as a \gls{daq} system. It is a system that actively influences (e.g., through samples) and returns the data in a digital format. This is in contrast to \emph{hardware components}, that passively affect the signals and data, e.g., an \gls{lna}. 
A testbed consists of one of several chains of \emph{data sources} and \emph{hardware components}.
An experiment can refer to an environment file, describing, e.g., the room dimensions, and 3D scans and models. 
The experiment includes the testbeds used and a number of measurements, each having different parameters. 

\subsubsection{Testbed Description Files}
A testbed description file is a hierarchy of logical chains of data sources and hardware components that reflect the real measurement/testbed setup. 
An example of a testbed description file, based on the Techtile testbed~\cite{9815696}, is depicted in \cref{fig:example-testbed-descr-file-techtile} and is based on the example shown in \cref{fig:testbed-example}. The testbed consists of two subsystems, an acoustic system, and an \gls{rf} infrastructure. The \gls{rf} data chain consists of \num{140} chains of an \gls{sdr} with, in this example, only one antenna connected to it (while two are possible). The acoustic part of the testbed encompasses one central \gls{daq} system connected, in this example, to \num{100} amplifiers and \gls{mems} microphones. Each \texttt{*} in \cref{fig:example-testbed-descr-file-techtile} represents a reference to the data source or hardware component, defined elsewhere (see the following sections).

\begin{figure}
    \centering
\begin{minted}[
frame=lines,
framesep=2mm,
baselinestretch=1,
fontsize=\scriptsize,
breaklines, mathescape
]
{yaml}
Techtile: &Techtile
  name: "Techtile"
  description: "Small description"
  url: "https://github.com/techtile-by-dramco"
  level: "L3" # Level of testbed, see $\text{\cite{callebaut20236g}}$
  data_chains:
    -   label: "RF"
        chain:
            data_source:
                <<: *B210
            channel_chain: 
                hardware_components: 
                    - <<: *Techtile915MHzAntenna
                data_source_channel: "0"
            num_data_source_chains: 140 # num_channels will be num_chains x 2 channels 
            channel_locations: # physical location of each channel source
                file: techtile_antenna_locations.npy 
                loc_unit: "m" # unit used for source location
    -   label: "Acoustic"
        chain:
            data_source:
                <<: *DAQ
            channel_chain: 
                hardware_components:
                    - <<: *TechtileMicrophonePA
                    - <<: *TechtileMicrophone
                data_source_channel: "0:100"
        num_data_source_chains: 1 # num_channels will be num_chains x 2 channels 
        channel_locations: # physical location of each channel source
            file: techtile_microphone_locations.npy 
            loc_unit: "m" # unit used for source location
\end{minted}
    \caption{Example of an incomplete testbed description file.}
    \label{fig:example-testbed-descr-file-techtile}
\end{figure}

\subsubsection{Data Source Description Files}

Each data source has a unique type associated with it, and each type in turn has a number of configurable parameters. 
An example of a data source is an \gls{sdr}, which has a set of configurable parameters such as, e.g., the sampling rate or bandwidth.

\subsubsection{Hardware Components}
Various hardware components, including antennas, cables, \glspl{pa}, filters, and sensors, are integral to testbeds and affect recorded datasets (see \cref{fig:testbed-example}). Knowledge of these systems is often required to calibrate hardware or to remove the effect of the antenna pattern when processing the dataset. Each hardware component can have one or several attributes associated with it. An example is presented in \cref{fig:testbed-example}, where the frequency response of the utilized amplifier is stored in a \gls{dss}-compliant dataset file.

\subsubsection{Environments}
Optionally, environment-related metadata can be stored, such as the physical properties of the room (air absorption, material properties, temperature, noise values, room dimensions, interference signals in surroundings, \dots). However, in \gls{dss} there are currently no constraints on the type of data and the storage format.

\subsubsection{Experiments}
The experiment description file contains the information about a conducted experiment and thus refers to other description files. The experiment description file can also contain variables such as temperature and references to photo and video files.

\subsection{Dataset}\label{sec:dataset}
The description files are utilized when reading/storing the data in a common format. 
A simple dataset structure is used to store the data of a specific experiment scenario. 
A tensor is used, which has a data and a time dimension.
These dimensions are tailored to the dataset type, exemplified by the type of channel measurement that is performed, e.g., \gls{rf} channel-sounding or acoustic measurements.
Next to predefined dataset types, a user of \gls{dss} is allowed to specify new types and thus extend the \gls{dss}.

The dataset is stored in an \gls{hdf5}~\cite{hdf5} or \gls{netcdf}~\cite{56302} format to maintain interoperability with a range of programming languages.

\subsection{Application Programming Interface}\label{sec:api}
The \gls{dss} \gls{api} functions as an additional layer atop the standardized storage and description files, enhancing the utility and adaptability of \gls{dss}. It is imperative to emphasize that, while the \gls{api} will be developed, \gls{dss} maintains its primary focus on the meticulous structuring of data and metadata.

The \gls{dss} \gls{api} is laid out to facilitate streamlined interaction with the standard and provide users with a versatile platform for extending functionalities. By adhering to the standardized data storage and metadata management conventions established by DSS, the API becomes a powerful tool for implementing common workflows and procedures. This approach allows \gls{dss} to become the unification of different libraries and tools for a wide range of measurement types. This way the community can focus on the experiments, and use already available post-processing procedures embedded in the \gls{dss} \gls{api}.

\section{Dataset Types -- DSS in the field}\label{sec:dataset-types}

\subsection{The Channel-Sounding Dataset}\label{sec:channel_sounding_dataset}
The dataset containing a channel-sounding experiment or scenario, see e.g.,~\cite{Payami2012, Martinez2014, Sommerkorn2019, Gaillot2021, Zelenbaba2021, Sandra2023, Sandra2022, Loeschenbrand2022, Stanko2023, WildingArxiv2023v1, 9539876}, has minimal three dimensions: the number of transmitters, the number of receivers, and the data in the delay or frequency domain. 
Each transmitter or receiver can have one or more channels, as specified by the testbed or data source description. 
The mapping between the transmit and receive channels should be described in the experiment description file. 
Furthermore, each channel can have a position and orientation associated with it, allowing to, e.g., take antenna characteristics into account. 
Similarly, data in the delay or frequency domain can have a delay or absolute frequency coordinate associated with it. 
Depending on the type of measurement hardware that is used, different means of (post-)processing can be necessary, such as reduction of the measurement bandwidth, selection of a specific frequency range for application scenarios, or pulse shaping.

Depending on the available data, examples for application of channel-sounding datasets are for testing of positioning or channel estimation algorithms (e.g., using array processing on suitable datasets), analysis of propagation characteristics for propagation/environment modeling or machine learning-based feature extraction, but also wireless power transfer applications, requiring as minimal knowledge the transmit power settings.
Below we briefly discuss the data that can be expected in a general channel-sounding dataset as well as two applications in form of positioning and wireless power transfer.

\subsubsection{Measurement Data}

\begin{figure}[t]
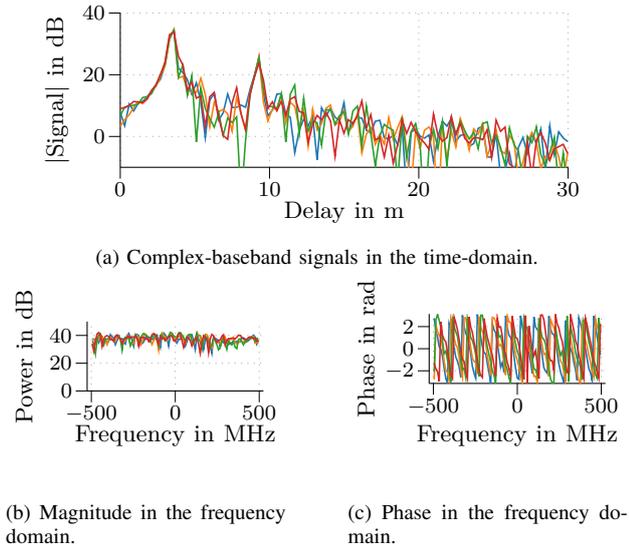

    \centering
    \begin{subfigure}[b]{\columnwidth}
         \centering
         \setlength{\figurewidth}{0.8\columnwidth}
        \setlength{\figureheight}{0.25\columnwidth}
         \input{figs/time_domain_log.tex}
         \caption{Complex-baseband signals in the time-domain.}
     \end{subfigure}
     
     \begin{subfigure}[b]{0.45\columnwidth}
         \centering
         \setlength{\figurewidth}{0.7\columnwidth}
        \setlength{\figureheight}{0.25\columnwidth}
         \input{figs/frequency_domain_mag.tex}
         \caption{Magnitude in the frequency domain.}
     \end{subfigure}\hfill
     \begin{subfigure}[b]{0.45\columnwidth}
         \centering
         \setlength{\figurewidth}{0.7\columnwidth}
        \setlength{\figureheight}{0.25\columnwidth}
         \input{figs/frequency_domain_phase.tex}
         \caption{Phase in the frequency domain.}
      \end{subfigure}

    \caption{Measurement signals between a single transmitter and $4$ exemplary receiver locations spaced roughly $2\,\mathrm{cm}$ apart.}%
    \label{fig:example:cr-tf}
\end{figure}

Visualization of the measurement data, e.g., the \gls{cir} between a specific transmitter $i_\mathrm{tx} = 0$ for each receive channel at a certain measurement time(index) $i_\mathrm{t}=0$ can be conveniently done via dedicated functions. 
These plot the channel response (CR), e.g., the received signal in complex baseband notation, or the complex-valued transfer function (TF) via
\begin{minted}{python}
dss.plot_cr((tx=0, rx=[0,1,2,3], t=0))
dss.plot_tf((tx=0, rx=[0,1,2,3], t=0))
\end{minted}
which allows to quickly view and compare measurements. Examples are shown in Fig.~\ref{fig:example:cr-tf} in form of the time- and frequency domain signals for a single transmitter location and $4$ receiver locations spaced roughly $2\,\mathrm{cm}$ apart for a measurement bandwidth of $1\,\mathrm{Ghz}$ and carrier frequency of $6.95\,\mathrm{GHz}$ (see \cite{WildingArxiv2023v1} for details regarding the measurements).

\subsubsection{Positioning}
For developing positioning applications, a multitude of algorithms can be employed, including various combinations of range- or angle-based algorithms or even \gls{rss}-based approaches. 
The data stored in \gls{dss} then needs to be formatted accordingly in post-processing to allow application of, e.g., channel estimation algorithms to extract ranges. 
This will commonly encompass the selection of the target bandwidth and center frequency, oversampling when required, or determination of parametric signal atoms for use with parametric channel estimators, see \cite[Sec.\,IV-C]{WildingArxiv2023v1}.
A corresponding \gls{dss}-compliant dataset will thus necessitate a certain level of calibration to be performed to enable correct use of the measurement data, ideally de-embedding the properties of the measurement device from the measurement data up to a minimum level that cannot be performed by the end-user, or would be required for algorithm testing.
This could encompass, e.g., calibration of the system response to remove effects that can be considered unique to the measurement hardware, but would generally require a description of the performed calibration procedure.
Positioning algorithms will furthermore require the location of all sensor positions, encompassing the \gls{ue}, agent or mobile device to localize, as well as the \glspl{bs}, often termed (positioning) anchors.
For the (positioning) anchors, information about the level of synchronization between these anchors is necessary. %
To incorporate hardware components such as antennas in algorithms, the orientation of these in the environment coordinate system needs to be specified in the \gls{dss} description file.

\subsubsection{Wireless Power transfer}

    Using a dataset to evaluate \gls{wpt} requires accurate amplitude measurements or simulations. 
    An essential part of the former is the calibration of linear systematic errors introduced by hardware components such as cables, antennas, connectors, etc. 
    A DSS-compliant dataset includes the characterization of these hardware components. The manufacturer of a data source guarantees calibrated measurements at its ports, and thus the measurement reference planes are located there. 
    
    It is evident that adopting DSS will ensure that applications working with channel-sounding data can seamlessly process data from different DSS-compliant datasets without particular care about the involved error networks.

\subsection{The Simulation Dataset}

In the multifaceted realm of 6G research, both real-world experimentation and theoretical modeling are paramount. Here, the simulation dataset stands out as a powerful instrument, providing researchers with a refined method to emulate scenarios, validate theories, and build models, even before initiating real-world experiments. The \gls{dss}, with its intricate architecture as described earlier, bestows upon the simulation dataset several noteworthy features.

At its foundation, the simulation dataset ensures dimensional rigor. Every simulation run represents an iteration or instance of a chosen scenario. In contrast, the associated result captures the observational outcome from that specific iteration. There's also an inherent fluidity in parameter selection. Simulations within the \gls{dss} framework often pivot around varying input parameters, such as network arrival rates. This variance unfolds a rich spectrum of results, from metrics like mean throughput and delay to volumes of dropped packets. Such depth of outcomes allows researchers to understand system behavior across different conditions.

Additionally, the importance of temporal granularity is hard to overstate. Many simulations seek to delineate how systems evolve over time. The \gls{dss} accommodates this by offering time-based outputs, structuring data points as time-value pairs, painting a clear picture of temporal variations. An experiment can generate a plethora of results. With the \gls{dss}, researchers can decide whether each system node should be a unique data source or if they should consider the entire simulation logging as a singular unit, which is crucial for nuanced data interpretation and analysis.

There are several overarching advantages to employing the simulation dataset within the \gls{dss}. Its unmatched flexibility allows researchers to control every variable, recreating myriad scenarios, some of which might be challenging or even impossible to stage in the real world. Economically, simulations often trump real-world experiments, especially when the latter require specialized or rare resources. The ability to rerun simulations guarantees reproducibility, reinforcing the credibility of findings. In scenarios where real-world tests might pose risks, simulations provide a hazard-free environment to venture into. Moreover, the inherent scalability of the \gls{dss} simulations ensures systems can be studied at any scale, from the smallest networks to sprawling infrastructures.

In essence, the simulation dataset, meticulously crafted within the \gls{dss} paradigm, is poised to become an indispensable tool in 6G network research. It bridges the theoretical with the practical, ensuring that researchers are equipped with robust, reliable, and insightful tools, preparing them for the demands of the real world.

\subsection{The Acoustic Dataset}
A dataset containing the \glspl{rir} or audio fragments obtained from different speakers and microphone combinations is desired, e.g.,~\cite{Glitza2023}. The acoustic dataset shows a close analogy to the channel-sounding dataset, described in~\cref{sec:channel_sounding_dataset}. The three dimensions for this dataset consist of the number of speakers, the number of microphones and the data (\glspl{rir} or audio fragments). The location dependency of a speaker or microphone entails the use of different channels, when focusing on array configurations. In simulations, it may be possible, for example, that single microphones are placed at all desired locations simultaneously, meaning that only 1 channel is applicable. Measurements are more likely to use and move the same microphone (array). Nevertheless, this dataset can be used for both simulation and measurements.

\begin{figure}[h]
    \centering
    \tikzsetnextfilename{RIR_example_pruned}
    \input{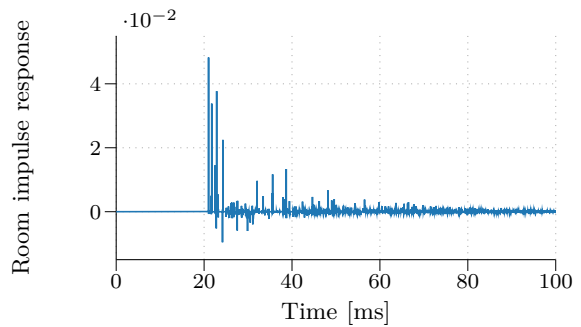}
    \caption{\Gls{rir} between the \nth{14} microphone and \nth{1} speaker.}
    \label{fig:example_RIR}
\end{figure}

Plotting the \gls{rir} of channel 0 between e.g., speaker~1 and microphone~14 can be conveniently done via:
\begin{minted}{python}
dss.plot_rir((sp=1, mic=14, ch=0))
\end{minted}
and results in, for example,~\cref{fig:example_RIR}.

\glsresetall
\section{Conclusion and future extensions}\label{sec:concl}
The \gls{dss} architecture presented in this document represents a robust framework for the structured storage and management of scientific datasets. \Gls{dss} offers a well-defined structure comprising description files, hardware component definitions, environments, and experiments, all working together to ensure the transparency, reusability, and interoperability of stored data. By focusing on standardized data storage and metadata management, \gls{dss} adheres to the \gls{fair} principles, making datasets findable, accessible, and usable for a wide range of scientific and engineering applications. The inclusion of a flexible API enhances \gls{dss}'s capabilities, enabling users to efficiently interact with and extend the standard for specific workflows and procedures. With its support for various dataset types, such as channel sounding and acoustic measurements, \gls{dss} proves versatile and adaptable to diverse research domains. In essence, the \gls{dss} architecture sets a strong foundation for effective data storage, retrieval, and analysis, contributing significantly to advancing scientific and experimental research through standardized and well-structured datasets.

The future outlook for \gls{dss} holds several promising opportunities for further development and extensions such as e.g., expanded dataset types, machine learning integration, and the extension of the API including data visualization tools.

In summary, the future of DSS is marked by its adaptability and openness to growth. By embracing emerging technologies and responding to the evolving needs of the scientific and engineering communities, DSS can continue to serve as a valuable standard for dataset storage and management, fostering greater collaboration and innovation in data-driven research.

\printbibliography

@article{56302,
	title        = {{NetCDF: an interface for scientific data access}},
	author       = {Rew, R. and Davis, G.},
	year         = 1990,
	journal      = {IEEE Computer Graphics and Applications},
	volume       = 10,
	number       = 4,
	pages        = {76--82},
	doi          = {10.1109/38.56302}
}

@article{9539876,
	title        = {{Experimental Exploration of Unlicensed Sub-GHz Massive MIMO for Massive Internet-of-Things}},
	author       = {Callebaut, Gilles and Willhammar, Sara and Guevara, Andrea P. and Johansson, Anders J. and Van Der Perre, Liesbet and Tufvesson, Fredrik},
	year         = 2021,
	journal      = {IEEE Open Journal of the Communications Society},
	volume       = 2,
	pages        = {2195--2204},
	doi          = {10.1109/OJCOMS.2021.3113088}
}

@inproceedings{9815696,
	title        = {{Techtile – Open 6G R\&D Testbed for Communication, Positioning, Sensing, WPT and Federated Learning}},
	author       = {Callebaut, Gilles and Mulders, Jarne Van and Ottoy, Geoffrey and Delabie, Daan and Cox, Bert and Stevens, Nobby and Perre, Liesbet Van der},
	year         = 2022,
	booktitle    = {2022 Joint European Conference on Networks and Communications \& 6G Summit (EuCNC/6G Summit)},
	pages        = {417--422},
	doi          = {10.1109/EuCNC/6GSummit54941.2022.9815696}
}

@misc{callebaut20236g,
	title        = {{6G Radio Testbeds: Requirements, Trends, and Approaches}},
	author       = {Gilles Callebaut and Liang Liu and Thomas Eriksson and Liesbet Van der Perre and Ove Edfors and Christian Fager},
	year         = 2023,
	eprint       = {2309.06911},
	archiveprefix = {arXiv},
	primaryclass = {eess.SP}
}

@software{DigitalRF,
	title        = {{Digital RF}},
	author       = {Volz, Ryan and Rideout, William C. and Swoboda, John and Vierinen, Juha P. and Lind, Frank D.},
	url          = {https://github.com/MITHaystack/digital\_rf},
	date         = {2022-12-07},
	version      = {2.6.8}
}

@software{dss2023,
	title        = {{Dataset Storage Interface}},
	author       = {Gilles Callebaut},
	year         = 2023,
	url          = {https://github.com/6G-Testbeds/Dataset-Storage-Interface}
}

@inproceedings{Gaillot2021,
	title        = {{Measurement of the V2I Massive Radio Channel with the MaMIMOSA Sounder in a Suburban Environment}},
	author       = {Gaillot, D.P. and Laly, P. and Dahmouni, N. and Delbarre, G. and Van den Bossche, M. and Vermeeren, G. and Tanghe, E. and Simon, E.P. and Joseph, W. and Martens, L. and Li\'{e}nard, M.},
	year         = 2021,
	month        = mar,
	booktitle    = {2021 15th {European} {Conference} on {Antennas} and {Propagation} ({EuCAP})},
	pages        = {1--4},
	doi          = {10.23919/EuCAP51087.2021.9410994}
}

@inproceedings{Glitza2023,
	title        = {{Database of Simulated Room Impulse Responses for Acoustic Sensor Networks Deployed in Complex Multi-Source Acoustic Environments}},
	author       = {Glitza, Rene and Becker, Luca and Martin, Rainer},
	year         = 2023,
	booktitle    = {31th European Signal Processing Conference (EUSIPCO)}
}

@article{gunst2018antenna,
	title        = {{Antenna data storage concept for phased array radio astronomical instruments}},
	author       = {Gunst, Andr{\'e} W and Kruithof, Gert H},
	year         = 2018,
	journal      = {Experimental Astronomy},
	publisher    = {Springer},
	volume       = 45,
	pages        = {351--362}
}

@online{hdf5,
	title        = {{Hierarchical Data Format, version 5}},
	author       = {{The HDF Group}},
	year         = {1997-NNNN}
}

@article{Loeschenbrand2022,
	title        = {{Towards Cell-Free Massive MIMO: A Measurement-Based Analysis}},
	author       = {David L{\"o}schenbrand and Markus Hofer and Laura Bernad\'{o} and Stefan Zelenbaba and Thomas Zemen},
	year         = 2022,
	journal      = {{IEEE} Access},
	publisher    = {IEEE},
	volume       = 10,
	pages        = {89232--89247},
	doi          = {10.1109/ACCESS.2022.3200365},
	issn         = {2169-3536},
	date         = 2022
}

@inproceedings{Martinez2014,
	title        = {{Towards very large aperture massive MIMO: A measurement based study}},
	shorttitle   = {Towards very large aperture massive {MIMO}},
	author       = {Mart\'{\i}nez, \`{A}lex Oliveras and De Carvalho, Elisabeth and Nielsen, Jesper \O{}dum},
	year         = 2014,
	month        = dec,
	booktitle    = {2014 {IEEE} {Globecom} {Workshops} ({GC} {Wkshps})},
	pages        = {281--286},
	doi          = {10.1109/GLOCOMW.2014.7063445},
	issn         = {2166-0077},
}

@software{NI-RF,
	title        = {{NI RF Data Recording API v1.0.0}},
	author       = {{National Instruments}},
	url          = {https://github.com/genesys-neu/ni-rf-data-recording-api}
}

@article{OnTheRoadTo6G,
	title        = {{On the Road to 6G: Visions, Requirements, Key Technologies, and Testbeds}},
	author       = {Wang, Cheng-Xiang and You, Xiaohu and Gao, Xiqi and Zhu, Xiuming and Li, Zixin and Zhang, Chuan and Wang, Haiming and Huang, Yongming and Chen, Yunfei and Haas, Harald and Thompson, John S. and Larsson, Erik G. and Renzo, Marco Di and Tong, Wen and Zhu, Peiying and Shen, Xuemin and Poor, H. Vincent and Hanzo, Lajos},
	year         = 2023,
	journal      = {IEEE Communications Surveys \& Tutorials},
	volume       = 25,
	number       = 2,
	pages        = {905--974},
	doi          = {10.1109/COMST.2023.3249835}
}

@inproceedings{Payami2012,
	title        = {{Channel measurements and analysis for very large array systems at 2.6 GHz}},
	author       = {Payami, Sohail and Tufvesson, Fredrik},
	year         = 2012,
	month        = mar,
	booktitle    = {2012 6th {European} {Conference} on {Antennas} and {Propagation} ({EUCAP})},
	pages        = {433--437},
	doi          = {10.1109/EuCAP.2012.6206345},
	issn         = {2164-3342}
}

@INPROCEEDINGS{Sandra2022,
  author={Sandra, Michiel and Nelson, Christian and Johansson, Anders J},
  booktitle={2022 IEEE Conference on Antenna Measurements and Applications (CAMA)}, 
  title={{Ultrawideband USRP-Based Channel Sounding Utilizing the RFNoC Framework}}, 
  year={2022},
  volume={},
  number={},
  pages={1-3},
  doi={10.1109/CAMA56352.2022.10002476}}

@article{Sandra2023,
	title        = {{Measurement-Based Wideband Maritime Channel Characterization}},
	author       = {Sandra, Michiel and Tian, Guoda and Fedorov, Aleksei and Cai, Xuesong and Johansson, Anders J},
	year         = 2023,
	journal      = {The 17th European Conference on Antennas and Propagation (EuCAP 2023)}
}

@software{sigMF,
	title        = {{SigMF Release v0.0.1}},
	author       = {Hilburn, Benjamin},
	year         = 2018,
	month        = jul,
	publisher    = {Zenodo},
	doi          = {10.5281/zenodo.1418396},
	url          = {https://doi.org/10.5281/zenodo.1418396},
	version      = {v0.0.1}
}

@inproceedings{Sommerkorn2019,
	title        = {{Experimental and Analytical Characterization of Time-Variant V2V Channels in a Highway Scenario}},
	author       = {Sommerkorn, Gerd and K\"{a}ske, Martin and Czaniera, Daniel and Schneider, Christian and Del Galdo, Giovanni and Thom\"{a}, Reiner S. and Walter, Michael},
	year         = 2019,
	month        = mar,
	booktitle    = {2019 13th {European} {Conference} on {Antennas} and {Propagation} ({EuCAP})},
	pages        = {1--5}
}

@inproceedings{Stanko2023,
	title        = {{Time Variant Directional Multi-Link Channel Sounding and Estimation for V2X}},
	author       = {Stanko, Daniel and D\"{o}bereiner, Michael and Sommerkorn, Gerd and Czaniera, Daniel and Andrich, Carsten and Schneider, Christian and Semper, Sebastian and Ihlow, Alexander and Landmann, Markus},
	year         = 2023,
	month        = jun,
	booktitle    = {2023 {IEEE} 97th {Vehicular} {Technology} {Conference} ({VTC2023}-{Spring})},
	pages        = {1--5},
	doi          = {10.1109/VTC2023-Spring57618.2023.10199213},
	issn         = {2577-2465}
}

@inproceedings{SurveyStandardsAstronomy,
	title        = {{A Survey on File Format for Data Storage in Radio Astronomy}},
	author       = {Chen, Meng and Deng, Hui and Wang, Feng and Ji, Kaifan},
	year         = 2013,
	booktitle    = {2013 6th International Conference on Intelligent Networks and Intelligent Systems (ICINIS)},
	pages        = {324--327},
	doi          = {10.1109/ICINIS.2013.90}
}

@misc{WildingArxiv2023v1,
	title        = {{Propagation Modeling for Physically Large Arrays: Measurements and Multipath Component Visibility}},
	author       = {Thomas Wilding and Benjamin J. B. Deutschmann and Christian Nelson and Xuhong Li and Fredrik Tufvesson and Klaus Witrisal},
	year         = 2023,
	archiveprefix = {arXiv},
	eprint       = {2305.05958},
	primaryclass = {eess.SP}
}

@article{wilkinson2016fair,
	title        = {{The FAIR Guiding Principles for scientific data management and stewardship}},
	author       = {Wilkinson, Mark D and Dumontier, Michel and Aalbersberg, IJsbrand Jan and Appleton, Gabrielle and Axton, Myles and Baak, Arie and Blomberg, Niklas and Boiten, Jan-Willem and da Silva Santos, Luiz Bonino and Bourne, Philip E and others},
	year         = 2016,
	journal      = {Scientific data},
	publisher    = {Nature Publishing Group},
	volume       = 3,
	number       = 1,
	pages        = {1--9}
}

@article{Zelenbaba2021,
	title        = {{Multi-Node Vehicular Wireless Channels: Measurements, Large Vehicle Modeling, and Hardware-in-the-Loop Evaluation}},
	shorttitle   = {Multi-{Node} {Vehicular} {Wireless} {Channels}},
	author       = {Zelenbaba, Stefan and Rainer, Benjamin and Hofer, Markus and L\"{o}schenbrand, David and Daki\'{c}, Anja and Bernad\'{o}, Laura and Zemen, Thomas},
	year         = 2021,
	journal      = {IEEE Access},
	volume       = 9,
	pages        = {112439--112453},
	doi          = {10.1109/ACCESS.2021.3100676},
	issn         = {2169-3536}
}

\end{document}